\documentstyle{article}
   \begin{document}
   \setlength{\baselineskip}{15pt}
   \title{Lotka-Volterra representation of general nonlinear systems}
   \author{ Benito Hern\'{a}ndez--Bermejo  \and
            V\'{\i}ctor Fair\'{e}n$^{*}$  }
   \date{}

   \maketitle
   {\em Departamento de F\'{\i}sica Fundamental,
   Universidad Nacional de Educaci\'{o}n a Distancia. Aptdo. 60141,
   28080 Madrid (Spain). \\
   E--mail: vfairen@uned.es}

\mbox{}

   \begin{center}
      {\bf ABSTRACT}
   \end{center}
In this paper we elaborate on the structure of the Generalized Lotka-Volterra
(GLV) form for nonlinear differential equations. We discuss here
the algebraic properties of the GLV family, such as the invariance under
quasimonomial transformations and the underlying structure of
classes of equivalence. Each class possesses a unique
representative under the classical quadratic Lotka-Volterra form. We show
how other standard modelling forms of biological interest, such as
S-systems or mass-action systems are naturally embedded into the
GLV form, which thus provides a formal framework for their comparison,
and for the establishment of transformation rules. We also focus
on the issue of recasting of general nonlinear systems into the GLV format.
We present a procedure for doing so, and point at possible sources of
ambiguity which could make the resulting Lotka-Volterra system dependent on
the path followed. We then provide some general theorems that define the
operational and algorithmic framework in which this is not the case.

\mbox{}

{\bf Running title:} Lotka-Volterra representation

\mbox{}

$^{*}$ Author tho whom all correspondence should be addressed.

\pagebreak
\section{Introduction}

\newtheorem{th}{THEOREM}
\newtheorem{co}[th]{COROLLARY}
\newtheorem{lm}[th]{LEMMA}
\newtheorem{pr}[th]{PROPOSITION}

The search and study of canonical representations (reference formats
which are form-invariant under a given set of transformations)
in nonlinear systems of ordinary
differential equations has been a recurrent theme in the literature. Although
the powerful algebraic structure which characterizes the theory of linear
differential systems does not seem to have a counterpart in the nonlinear
realm (a not so surprising fact, once we take into consideration
the apparent diversity in structure and richness of behaviors
of nonlinear vector fields), there is an increasing number of suggestions
for a partial solution to this problem, which we shall partly review
later in this paper.

Recasting an n-dimensional differential system into a canonical form conveys
a gain in algebraic order which is not without cost, for it has usually
to be embedded in a higher dimensional mathematical structure. The
procedure is then justified if in the context of the target canonical form
we are in possession of powerful mathematical tools allowing for a better
analysis of the original system. This is not always the case, for not all
suggested canonical forms provide, in this sense, a satisfactory level.

The Lotka-Volterra structure can be considered one of the favored forms
to this effect. First, it may qualify for canonical form in
a classical context, for Plank \cite{pla} has demonstrated
that $n$-dimensional Lotka-Volterra equations are hamiltonian, and are thereby
amenable to a classical canonical description once an appropriate Poisson
structure is chosen. Second,
its paramount importance in ecological modeling equals
its ubiquity in all fields of science, from plasma physics \cite{lav} to
neural nets \cite{noo}. This may not be unrelated to the fact that it is quadratic,
and thus appears in many models in which interaction processes are
viewed as  fortuitous `collisions', or `encounters', between at most two
constitutive entities;
those with more than two participants being seen as extremely unprobable.
Additionally, it is characterized as simple algebraic objects as matrices,
which makes its analysis far more attractive than that of other formats.
Also, being representable in terms of a network, it spans a bridge to a
possible connection to a graph theory approach in the qualitative
study of nonlinear differential equations, either directly
or through its equivalence with the well known replicator
equations \cite{hys}.

The purpose of this article is to elaborate on the algebraic structure
of the so called GLV formalism, defined on equations the structure of
which generalizes the $n$-dimensional Lotka-Volterra system -it
contains them as a particular case-. It thus offers a natural bridge
towards the representation of nonlinear systems in terms of Lotka-Volterra
equations. We will also review
other known canonical forms, paying special attention, due to its biological
implications, to the so called S-system format, introduced by Savageau and
coworkers as a potential way of approaching nonlinear systems
(see Savageau, Chap.1 in \cite{vo1}). We will show
how the GLV formalism offers a formal solution to the issue of
transformations between different canonical forms, a problem which has
already attracted the attention of Savageau and Voit in the case involving
Lotka-Volterra and S-system forms \cite{vo2}.

Despite the versatility of the GLV equations,
they do not seem {\em at first sight} to encompass many
model systems of biological or physical significance, with, for example,
saturating rates defined in terms of rational functions.
This exclusion is, however, only apparent, for
it is a well kown fact that non-polynomial rate laws are always amenable
to a polynomial format by the introduction of conveniently chosen
auxiliary variables. This trick, which was known from old in the field of
Celestial Mechanics \cite{sie}, was popularized by Kerner
\cite{ker}, and independently by Savageau and
collaborators (see Voit, Chap.12 in \cite{vo1}),
who have made ample use of this technique. Although the GLV equations are
somewhat different from polynomial ones, there is no obstacle for aplying
the same procedure, as it will be shown. The problem is that
the above technique is unfortunately not systematic, as
it relies on a clever selection of the additional variables
and of their derivatives (as we shall see later) and does not generally
lead to a unique system in anyone of the desired formats, let it be
polynomial, S-system, or any other whatsoever (in our case in GLV form).
This ambiguity, and the resulting multiplicity of target systems (an infinite
number is not so uncommon), may throw some shadow on the procedure, and be
especially confusing when a single choice of auxiliary variables leads
to the disclosure of several entirely different target systems, or reversely,
when a single target system originates from completely different choices of
auxiliary variables.

No doubt, the previous method of auxiliary variables has important drawbacks,
but it is presently the only known course of action when confronted to this
type of recasting problem. The limits of these ambiguities should be then
clearly outlined, for it is essential to enhance the confidence in the
applicability of the method. This task is carried out in the final sections
of the article.

\section{Overview of canonical forms}

{\sc A) Infinite-dimensional linear systems.}
If the theory of linear vector fields has been given a well defined
and coherent structure, can we somehow linearize? This sensible question
was given an answer in 1931 by T. Carleman \cite{car},
following Poincar\'{e}'s suggestion. He showed that a finite-dimensional
system of ordinary polynomial differential equations
is equivalent to an infinite-dimensional linear system of ODE's.
The whole issue lay dormant until the late seventies.
Since, several authors
have greatly contributed to the investigation of the potential applications
of the Carleman embedding, which have been recently reviewed by
Kowalski and Steeb \cite{kys}.
Although there has been an interesting suggestion of a `quantum
mechanical' formalism applicable to the Carleman linearization,
the scheme does not seem to provide, for the time being, 
an operationally acceptable framework. Additionally,
the manipulation of an infinite-dimensional system, as linear as it
may be, can still be considered objectable by many users.

{\sc B) Riccati systems.}
Some time ago Kerner \cite{ker} proposed a scheme with the purpose of
bringing general nonlinear differential systems down to
polynomial vector fields, and ultimately to what he termed
{\em elemental Riccati systems\/}:
\[
\dot x_i = \sum_{j , k} A_{i}^{jk} x_j x_k,
\]
with $A_{i}^{jk}$ either 0 or 1. He suggests to do so
by introducing step by step new variables which represent collectives of
other variables and, by
making use of the sequential differentiation rule -in a way similar
to that of the Carleman embedding-.
This rather heuristic recipe is the traditional, and widely used,
method for reducing the degree of a nonlinearity. The dimension of the
elemental Riccati system is, of course, greater than that of the initial
system; but the gain in structure -so to say, in order- is not costless.

{\sc C) Mass action systems.}
Chemical kinetics has been considered by certain authors a good candidate for
prototype in nonlinear science \cite{eyt}.
They claim that it would already deserve this
consideration if it were only because
it embraces all types of behavior of interest, from multiplicity
of steady states to chaotic evolution, with the backing of a large
{\em corpus} of experimental evidence. The simplicity
of the stoichiometric rules and that of the algebraic structure
of the corresponding evolution equations has made chemical
kinetics a traditional point of reference in modeling
within such fields as population biology \cite{pie},
quantitative sociology \cite{wei}, prebiotic evolution \cite{hys} and other
biomathematic problems \cite{mur}, where a system is viewed as a collection
of `species' interacting as molecules do.
As emphazised by Erdi and T\'{o}th \cite{eyt}, even the algebraic
structure of the evolution equations from many other fields
can be converted into  `chemical language', where a formal `analog'
in terms of a chemical reaction network is defined.
However, the serious obstacle of negative cross-effects was emphasized by
T\'{o}th and H\'{a}rs \cite{tyh}, by showing that no orthogonal transformation
leads the Lorenz and R\"ossler systems to a `kinetic' format.
Although many suggestions have been made in order to overcome the difficulty
of the negative cross-effects \cite{kow,pol,sgw,vo3}, the problem seems to
remain unsolved.

{\sc D) S-systems.}
S-systems constitute an interesting canonical form that has been developed
in the context of the power-law formalism in theoretical biochemistry. Its
proponents have made a considerable effort in showing how it
is a good candidate for representing general nonlinear systems, as
well as in elaborating on its relation to other forms, from
generalized mass-action to Lotka-Volterra systems
(See Voit, Chap. 12 in \cite{vo1}).
Its particularly simple form
\[
  \dot{x}_{i} = \alpha _{i}\prod_{j = 1}^{n} x_{j}^{g_{ij}} - \beta
   _{i}\prod_{j = 1}^{n} x_{j}^{h_{ij}} , \;\: i = 1, \ldots, n \; \: ,
\]
optimal estimation of parameter values from steady-state experiments
and the possibility of symbolic steady-state analysis
(See Weinberger, Chap. 6 in \cite{vo1})
have been proposed, among others, as arguments to justify the choice of
S-systems. Although some preliminary steps have been covered
(see Voit, Chap. 15 in \cite{vo1}, and also \cite{vof}),
much work is still necessary to provide the S-systems formalism
with a proper formal framework yielding a workable algebraic structure,
wherefrom insight on their
mathematical properties might be gained.
We will pay special attention to S-systems in the present paper. We will
do it by showing how they find their place within the
generalized Lotka-Volterra formalism.

{\sc E) Lotka-Volterra systems.}
The well-known $n$-dimensional Lotka-Volterra (LV) equations,
\begin{equation}
\dot{x}_i = \lambda_i x_i + x_i \sum_{j=1}^n A_{ij}x_j, \;\: i=1,...,n,
\label{kklv}
\end{equation}
have obvious quadratic nonlinearities and are characterized by simple
algebraic objects: matrices $\lambda$ and $A$.
Though they have occupied a
priviledged position in ecology -practically all high dimensional {\em
strategic} models are set in terms of them- they also appear in many other
fields, such as virology, where the concept of {\em quasispecies}
has given
a whole new perspective \cite{eyc,ewo}.
Cair\'{o} and Feix \cite{cyf} refer to a fairly long list of systems modeled
by LV equations; a sample which speaks in favor of their representative role,
and that has prompted Peschel and Mende \cite{pym} to head their book
on the issue with the title: {\em Do we live in a Volterra World?}
We may also recall that Lotka-Volterra equations are also equivalent to game
dynamical equations, replicator or autocatalytic networks \cite{hys}.
Through this connection, LV dynamics is linked to the whole fruitful
field of replicator dynamics and autocatalytic networks,
which is a continuous source of modeling in prebiotic evolution,
game dynamics, or population genetics.
LV systems will be given a priviledged status in what is to follow.

\section{Generalized Lotka-Volterra formalism}

The term {\em generalized Lotka-Volterra equations} (GLV) has been recently
coined by Brenig \cite{bre1} to refer to a system of the following form:
\begin{equation}
\dot{x}_i = \lambda_i x_i + x_i \sum_{j=1}^m A_{ij}
\prod_{k=1}^n x_k^{B_{jk}}, \;\: i=1,...,n,
\label{4}
\end{equation}
where $m$ is a positive integer not necessarily equal to $n$. Following Brenig
\cite{bre1}, vectors {\bf x} and $\lambda$, and the $n\times m$ matrix $A$
and $m\times n$ matrix $B$ may be indifferently real or complex. However, we
shall assume in what follows that the $x_i$ are real and positive and that
the matrix entries are arbitrary real numbers. The importance of (\ref{4}) as
a representation of Lotka-Volterra models was previously studied by Peschel
and Mende (see \cite[p. 120 ff]{pym}), who anticipated many of the interesting
properties of the algebraic structure of (\ref{4}) (they termed them
{\em multinomial differential systems}). These equations also appear in
independent developments by Br'uno \cite{bru} and Gouz\'e \cite{gou}. They
embrace a large category of relevant systems of differential equations, and
can be considered as equivalent to the Generalized Mass Action systems (GMA)
which have been dealt with by Savageau and coworkers \cite{vo1}.

Several important properties reveal the potential interest of the GLV equations
(\ref{4}). We may start by recalling some propositions from Peschel and Mende
\cite[Sec. 5.2]{pym}, Brenig and Goriely \cite{bre2} and Hern\'{a}ndez--Bermejo
and Fair\'{e}n \cite{bhyvf}, which we summarize in a single Theorem:
\begin{th} \label{thg}
\mbox{}  \\
i) GLV equations (\ref{4}) are form-invariant under quasimonomial power
transformations:
\begin{equation}
x_i = \prod_{k=1}^n y_k^{C_{ik}},\;\: i=1,...,n,
\label{5}
\end{equation}
defined by any non-singular (in our case, real) $n\times n$ matrix, C. In
other words, the system of equations obtained from eqs. (\ref{4}) by a
quasimonomial transformation, (\ref{5}), is also a GLV system of
the same dimension. Moreover, if we denote such system as
\begin{equation}
\dot y_i = \hat{\lambda}_i y_i + y_i \sum_{j=1}^m \hat{A}_{ij}
\prod_{k=1}^n y_k^{\hat{B}_{jk}},\;\: i=1,...,n,
\label{6}
\end{equation}
then:
\begin{equation}
\hat{\lambda}=  C^{-1} \cdot \lambda, \;\: \hat{A}= C^{-1} \cdot A,
\;\: \hat{B} = B \cdot C \;\: .
\label{7}
\end{equation}

ii) The product matrices,
\begin{equation}
\hat{B} \cdot \hat{\lambda}= B \cdot \lambda, \;\:
\hat{B} \cdot \hat{A} = B \cdot A ,
\label{8}
\end{equation}
are invariants under the quasimonomial transformations (\ref{5}). The
whole family of systems (\ref{4}) is then split into
classes of equivalence according to
relations (\ref{8}), such that, for given values of n and m,
to each class of equivalence
specific realizations of the product matrices $B \cdot \lambda$ and
$B \cdot A$ can be associated.

iii) The quasimonomials
\begin{equation}
\prod_{k=1}^n x_k^{B_{jk}}, \;\: j=1,...,m
\label{9}
\end{equation}
constitute a set of m invariants of the class of equivalence to
which the corresponding GLV system belongs.

iv) All GLV systems (\ref{4}) defined in an open subset of the positive
orthant which belong to the same class of equivalence are topologically
equivalent, that is, their phase spaces can be mapped into each other by a
diffeomorphism \cite{jac}, given by (\ref{5}).
\end{th}

In particular, the importance of quasimonomial transformations in what follows
cannot be underestimated. Their relevance has been clearly emphasized in the
literature (see \cite[Secs. 5.2 and 5.4]{pym} and \cite{bre1,bru}). These
transformations have been also used by Voit to study symmetry properties of
GMA systems in \cite[Ch. 15]{vo1} and \cite{vof}.

\subsection{The Lotka-Volterra canonical form}

In order to go further ahead in detailing the features of the GLV formalism
in the context of its canonical forms we should now distinguish three
independent cases, two of which have been studied by Brenig and Goriely
($m=n$, $m>n$) while the third ($m<n$) is
considered in here for the first time. We shall find necessary to elaborate
on them all, for they will help us in understanding the recasting technique
which we shall later on use for embedding S-systems into the GLV formalism.

\subsubsection{Case $m = n$}

$A$ and $B$ in (\ref{4}) are $n\times n$ square matrices. We
consider some specific transformation matrices $C$ which lead to interesting
canonical forms. Assume first that $B$ is invertible and $C$ is taken as
$B^{-1}$. According to (\ref{7}) $\hat{B}$ reduces to the identity matrix and
(\ref{6}) takes the usual LV form,
\begin{equation}
\dot y_i = \hat{\lambda}_i y_i + y_i \sum_{j=1}^n \hat{A}_{ij} y_j,
\;\: i=1,...,n,
\label{10}
\end{equation}
By construction, for
those classes of equivalence with non-singular matrices $B$
there is a unique LV representative.
It is interesting to observe that, according to (\ref{5}), each of the
variables $y_j$ in (\ref{10}) is actually
\begin{equation}
y_j = \prod_{k=1}^n x_k^{B_{jk}},\;\: j=1,...,n.
\end{equation}
In other words, each of the variables in the LV scheme (\ref{10})
accounts for one of the different nonlinear quasimonomials in (\ref{4}).

\subsubsection{Example with $m = n$:}
We shall reduce to the Lotka-Volterra canonical form the generic GLV system:
\begin{eqnarray*}
   \dot{x}_1 & = & x_1 [ \lambda_1 + a_{11}x_1^p + a_{12}x_2^q ]  \\
   \dot{x}_2 & = & x_2 [ \lambda_2 + a_{21}x_1^p + a_{22}x_2^q ]
\end{eqnarray*}
If we perform a transformation of the form (\ref{5}) with matrix
$C = B^{-1}$, we are led to a LV system with matrices $ \hat{A} = B \cdot A $
and $\hat{\lambda} = B \cdot \lambda $. It is:
\begin{eqnarray*}
   \dot{y}_1 & = & y_1 [p\lambda_1 + pa_{11}y_1 + pa_{12}y_2 ] \\
   \dot{y}_2 & = & y_2 [q\lambda_2 + qa_{21}y_1 + qa_{22}y_2 ] \; ,
\end{eqnarray*}
where $y_1 = x_1^p$ and $y_2 = x_2^q$.

\subsubsection{Case $m > n$}

Here, the number of quasimonomials $m$ is higher than that of independent
variables. Accordingly, the target LV
form (\ref{10}) is to be an $m$-dimensional system, its variables
standing for the $m$ quasimonomials in (\ref{4}). The transformation of
section 3.1.1 cannot be carried out unless (\ref{4}) is previously
embedded in an equivalent $m$-dimensional system.
To do so we enlarge system (\ref{4})
by introducing $m-n$ auxiliary `arguments', to which we assign
a fixed value, $x_{l}= 1, \l=n+1,...,m$, and that enter the equations in the
following way:
\begin{equation}
\dot x_i = \lambda_i x_i + x_i \sum_{j=1}^m A_{ij}
\prod_{k=1}^n x_k^{B_{jk}}\cdot [x_{n+1}^{B_{j,n+1}}\ldots x_{m}^{B_{jm}}],
\;\: i=1,...,n,
\label{15}
\end{equation}
with arbitrary values of $B_{j,n+1},...,B_{jm}$:
we are in fact adding $m-n$ arbitrary columns
to the $m\times n$ matrix $B$ in order to complete a non-singular
$m\times m$ matrix $\tilde{B}$.
In (\ref{15}), the term
in brackets should not affect the equations as long as the new
arguments stick to their assigned value. We do ensure it by defining
for them the equations:
\begin{equation}
\dot x_{l} = \lambda_{l} x_{l} + x_{l} \sum_{j=1}^m A_{l j}
\prod_{k=1}^n x_k^{B_{jk}}\cdot [x_{n+1}^{B_{j,n+1}}\ldots x_{m}^{B_{jm}}],
\;\: l=n+1,...,m,
\label{16}
\end{equation}
with entries $\lambda_{l}=0$ and $A_{l j}=0$, for $l=n+1,...,m$,
and initial conditions, $x_{l}(0)=1$.
Then, (\ref{15}) and (\ref{16}) define an expanded $m$-dimensional system to
which the procedure of subsection 3.1.1 can be applied. This embedding
technique preserves the topological equivalence between the initial and
final systems, as has been demonstrated in \cite{bhyvf}.

\subsubsection{Example with $m > n$:}

As an example, we shall consider a simple spheroid-model for tumor growth,
due to Maru\v{s}i\'{c} {\em et al.\/} \cite{marus}:
\[
    \dot{V} = V[-3\omega + 3\alpha k^{1/3} V^{-1/3} -3\alpha k^{2/3} V^{-2/3}
               +\alpha k V^{-1}], \;\:\; \alpha, k, \omega >0.
\]
Here $V$ denotes the tumor volume, provided $V \geq k$. We perform the
embedding described by equations (\ref{15}) and (\ref{16}). The matrices of
the expanded system are given by:
\[
  \tilde{A} =
      \left( \begin{array}{ccc}
      3\alpha k^{1/3} & -3\alpha k^{2/3} & \alpha k \\
      0 & 0 & 0 \\
      0 & 0 & 0
   \end{array} \right) \: ; \;\:
  \tilde{B} = \left( \begin{array}{ccc}
      -1/3 & 0 & 0 \\ -2/3 & 1 & 0 \\ -1 & 0 & 1
      \end{array} \right) \: ; \;\:
  \tilde{\lambda} = \left( \begin{array}{c} -3\omega \\ 0 \\ 0 \end{array} \right)
\]
We are now in the case $n = m$. The resulting LV system is thus:
\begin{eqnarray*}
  \dot{y}_1 & = & y_1 [\omega + \mu_1 y_1 + \mu_2 y_2 + \mu_3 y_3] \\
  \dot{y}_2 & = & y_2 [2 \omega + 2 \mu_1 y_1 + 2 \mu_2 y_2 + 2 \mu_3 y_3] \\
  \dot{y}_3 & = & y_3 [3 \omega + 3 \mu_1 y_1 + 3 \mu_2 y_2 + 3 \mu_3 y_3] \; ,
\end{eqnarray*}
where $\mu_1 = -\alpha k^{1/3}$, $\mu_2 = \alpha k^{2/3}$ and $\mu_3 =
-\alpha k/3$.

\subsubsection{Case $m<n$}

In this case, the number of quasimonomials, $m$, is smaller than that of
variables, $n$.
Consequently, there is no need to perform an embedding, as in the previous
case. Only $m$ variables of the n-dimensional LV system will
stand for the $m$ original quasimonomials, while the $n-m$ remaining
variables of that same LV system, as we shall see, will have an arbitrary
dependence on the original
variables. This means, as we can guess, that only $m$ variables are actually
independent. In fact, we demand to the $m \times n$ $\hat{B}$ matrix of the
target LV system to be of the form $\hat{B} = ( I_{m \times m} \mid
0_{m \times (n-m)} )$ (save row and column permutations), where $I$ is the
identity matrix, $0$ is the null matrix, and the subindexes indicate the
sizes of these submatrices. On the other hand, we also have from (\ref{7}),
$\hat{B}=B \cdot C$. If $Z$ denotes the inverse of matrix $C$, we have
$\hat{B} \cdot Z = B$. Since the structure of $\hat{B}$ is very simple, we
can explicitly evaluate, and write:
\begin{equation}
   B =
   \left( \begin{array}{cccc}
      z_{11} & \ldots & z_{1n} \\
      \vdots & \ddots & \vdots \\
      z_{m1} & \ldots & z_{mn}  \end{array} \right)
   \label{corte}
\end{equation}
Thus, the first $m$ rows of $Z = C^{-1}$ are given by the entries of matrix
$B$ from the original GLV system. Since $C$ must be an invertible matrix, we
have demonstrated the following result:
\begin{th}
\mbox{}  \\
If $m<n$, a necessary and sufficient condition for the existence of a
transformation (\ref{5}) leading to a LV system is that matrix B is of
rank m.
\end{th}
The procedure continues straightforwardly by completing the $n-m$ undefined
rows of $Z$ with arbitrary selected entries which make the resulting matrix
invertible. There are infinite ways of doing
this, generating a multiplicity of LV systems present in
the corresponding class of equivalence. If $m=n$, we would obtain directly
$Z = B$: a known result from 3.1.1. In other words, all features of the
limit case $n=m$ are preserved when $m<n$, with the only exception of the
uniqueness of the LV system: now we have infinite LV systems in a given
class of equivalence.

\subsubsection{Example with $m < n$:}
We shall consider the following GLV system:
\begin{eqnarray}
\dot{x}_{1} & = & x_{1}(-1 +  x_{1}x_{2}x_{3} +  x_{1}^{2}x_{2}x_{3}^{2} ) \label{prin} \\
\dot{x}_{2} & = & x_{2}( 3 + 2x_{1}x_{2}x_{3} +  x_{1}^{2}x_{2}x_{3}^{2} ) \\
\dot{x}_{3} & = & x_{3}( 2 -  x_{1}x_{2}x_{3} + 2x_{1}^{2}x_{2}x_{3}^{2} ) \label{fin}
\end{eqnarray}
We choose the LV matrix $B$ of the form (the row permutation is irrelevant):
\[  \hat{B} = \left( \begin{array}{ccc}
               0 & 1 & 0 \\ 1 & 0 & 0 \end{array} \right) \]
Consequently:
\[   \left( \begin{array}{ccc}
      z_{21} & z_{22} & z_{23} \\ z_{11} & z_{12} & z_{13} \end{array} \right) = B =
      \left( \begin{array}{ccc}
      1 & 1 & 1 \\ 2 & 1 & 2 \end{array} \right) \]
We now complete $Z$ with an arbitrary third row: $(1 \; 0 \; 0)$ for example.
Then $ C = Z^{-1} $. After evaluation of $C$, $\hat{A}$ and $\hat{\lambda}$
can be obtained, and the resulting LV system is:
\begin{eqnarray*}
\dot{y}_{1} & = & y_{1}( 5 + 7y_{1} + 2y_{2}) \\
\dot{y}_{2} & = & y_{2}( 4 + 4y_{1} + 2y_{2}) \\
\dot{y}_{3} & = & y_{3}(-1 +  y_{1} +  y_{2})
\end{eqnarray*}
where $y_{1} = x_{1}^{2}x_{2}x_{3}^{2}$ and $y_{2} = x_{1}x_{2}x_{3}$, and
stand for the two quasimonomials present in (\ref{prin})--(\ref{fin}).
As mentioned above, the $n-m$ remaining variables (here $y_{3}$) have been
freely chosen, a fact that implies the selection of one of the infinite
existing LV systems in the class of equivalence. We should nevertheless,
notice that the $m$ actually independent variables (here $y_1$ and $y_2$)
obey a unique Lotka-Volterra system. As far as this is valid we can also
claim in this case the uniqueness of the Lotka-Volterra as representative of
a class of equivalence.

\subsection{Single-quasimonomial canonical form}

We shall briefly mention another form which will prove to be of interest
in later sections (see \cite[p. 124]{pym}).

\subsubsection{Case $m=n$}

For any class of equivalence for which $A$ is non-singular
we can choose $C=A$. In which case we have for (\ref{6}),
\begin{equation}
\dot y_i = \hat{\lambda}_i y_i + y_i
\prod_{k=1}^n y_k^{\hat{B}_{ik}},\;\: i=1,...,n.
\label{14}
\end{equation}
This canonical form has interesting integrability properties
which have been studied in depth by Goriely and Brenig \cite{gori1}.

On the other hand,
it is clear that if all
$\lambda$'s are nonpositive, equation (\ref{14}) is also an S-system.
We shall find this similarity useful and
consider in a forthcoming section an example
of the single quasimonomial canonical form under the viewpoint of the
S-system recasting technique.

\subsubsection{Case $m>n$}

As done in subsection 3.1.3,
a preliminary embedding of the original GLV equation
in a m-dimensional
system is also a prescriptive requirement prior to the application
of the procedure for case $m=n$ and the obtainment of this
canonical form. However, this time the embedding must
be such
that the extended matrix $A$ be invertible.
In order to satisfy this condition, Brenig and Goriely \cite{bre2} introduce
$m-n$ auxiliary variables such that:
\[
  \dot{x}_{\alpha} = \left\{ \begin{array}{l}
  \lambda _{\alpha} x_{\alpha} + x_{\alpha} \sum_{\beta=1}^{m} A_{\alpha \beta} \prod_{\gamma=1}^{n}
   x_{\gamma}^{B_{\beta \gamma}} \cdot [x_{n+1}^{0} \ldots x_{m}^{0}] , \;\:\; \alpha = 1, \ldots , n  \\
   \rho _{\alpha} x_{\alpha} + x_{\alpha} \sum_{\beta=1}^{m} a_{\alpha \beta} \prod_{\gamma=1}^{n}
   x_{\gamma}^{B_{\beta \gamma}} \cdot [x_{n+1}^{0} \ldots x_{m}^{0}] ,\;\:\; \alpha = n+1, \ldots ,  m
   \end{array} \right.
\]
We can then proceed as in subsection 3.2.1. A proof, which is rather involved,
of the topological invariance of the solutions under this embedding process
will be provided in a future work.

\section{S-systems within the GLV formalism}

The observation that a generic S-system
\[
  \dot{x}_{i} = \alpha _{i}\prod_{j = 1}^{n} x_{j}^{g_{ij}} -
       \beta _{i}\prod_{j = 1}^{n} x_{j}^{h_{ij}} , \;\: i = 1, \ldots, n,
\]
is in fact a particular case of GLV system, allows us to focus the attention
to applying the previous formalism to the inverse problem, that is, the
conversion of a GLV system into an equivalent S-system. We shall see how the
recasting procedure can now be standarized within the formalism we have
presented. We complement the multiple heuristic recipes quoted for this
purpose (see Voit, Ch.12 in \cite{vo1}) by presenting a formal strategy
which allows a non-negligeable freedom of
design of target S--system matrices, within the obvious
restriction of topological equivalence (TE). This TE will be ensured if
the target S--system is reachable through any element of the infinite
group of quasimonomial transformations. This, together with the ability to
predict the dimension, number and exact definition of the variables are some
of the advantages derived from dealing with the problem in a well
characterized mathematical framework.

We shall again study the problem of recasting into S--systems in three steps
of increasing complexity:

\subsection{Case $m=n$}

\subsubsection{Recasting theory}

We start from a GLV system:
\begin{equation}
   \dot{x_{i}} = x_{i}(\lambda _{i} + \sum_{j=1}^{n}A_{ij}\prod_{k=
      1}^{n}x_{k}^{B_{jk}}) , \;\: i = 1 \ldots n .
   \label{eq:glv2}
\end{equation}
We must apply a quasimonomial transformation to (\ref{eq:glv2}) and proceed 
consequently to a new GLV system (with matrices $\hat{\lambda}$, $\hat{A}$, 
$\hat{B}$) {\em to which we also demand the fulfillment of an S--system 
format specifications.\/} This target system must be designed taking into 
account the fact that, according to the S--system form, in every equation 
there is only one positive and one negative term (the linear term is also a 
possible term in an S--system). In order to find the right transformation 
matrix, $C$, we rewrite equations (\ref{7}) by introducing the {\em extended 
matrices E} and $D$, associated to the GLV system (\ref{eq:glv2}) and to the 
transformation matrix $C$ respectively:
\begin{equation}
  E = \left( \begin{array}{cc}
      1        & \vec{0}^{\; t}  \\
      \lambda  &  A
      \end{array} \right) \;\: , \;\:
  D = \left( \begin{array}{cc}
      1        & \vec{0}^{\; t}  \\
      \vec{0}  &  C
      \end{array} \right) \;\: ,
      \;\;\: \{ E , D \} \subset {\cal M}_{(n+1)\times(n+1)} \: .
  \label{nm2}
\end{equation}
Then the GLV system matrices $A$, $B$ and $\lambda$ are equivalently
specified by $E$ and $B$. The equivalence between $C$ and $D$ is obvious.
When a transformation of the kind (\ref{5}) is performed, the GLV matrices
change to:
\begin{equation}
\hat{B}=B \cdot C, \;\: \hat{E} = D^{-1} \cdot E
     \label{nm3}
\end{equation}
By means of this construction, which will be very useful later,
the following result can be easily demonstrated:

\begin{th}  \label{th7}
\mbox{} \\
Suppose that $m = n$ and that matrices $A$ and $\lambda$ of system
(\ref{eq:glv2}) possess arbitrary real entries. Assume the target
S--system is characterized by some specifically chosen matrices
$\hat{\lambda}$ and $\hat{A}$.
Then, there exists a unique transformation of the form (\ref{5}),
given by $C = A \cdot \hat{A}^{-1}$, which leads
from (\ref{eq:glv2}) to such S--system, provided matrices
$A$ and $\hat{A}$ are regular and vectors $\lambda$
and $\hat{\lambda}$ can be related by the compatibility condition
 $\lambda = A \cdot \hat{A}^{-1} \cdot \hat{\lambda} $.
\end{th}

Notice also that in the general case the design of the final S--system can be
done easily working directly with the extended matrix $\hat{E}$. That is, the
rule is that every row of $\hat{E}$ must contain only two nonzero elements,
one of which is positive and the other negative.

On the other hand, the compatibility condition of Theorem \ref{th7} yields a system of
equations with $n^{2}$ unknown quantities (the entries of matrix $C = A \cdot
\hat{A}^{-1}$). The application of the Rouch\'{e}--Fr\"{o}benius theorem
\cite{lip}
shows that there will always be infinite solutions to such a system, thus
ensuring precisely the fulfillment of that same compatibility condition.

As we can see, there exist infinite different S--systems in every class of
equivalence.
In the case $n=m$, however, there is a single LV system, provided matrix
$B$ is regular: the one resulting through the election $C = B^{-1}$. In fact,
it is straighforward to notice that in this case such LV system can be chosen
as the canonical element of the class.

A remarkable feature of the previous theory is the complete freedom in the
choice of the form for matrix $\hat{A}$. This means that the last theorem
can be applied equally to any kind of canonical form, not necessarily that of
an S--system. For example, choosing a diagonal matrix $\hat{A}$ we construct
a family of systems which includes the single quasimonomial canonical form as
a special case \cite{bre1,bre2}. As we shall see subsequently, this freedom
is maintained in the other general situations.

\subsubsection{Example with $m=n$.}
We shall recast the system of section 3.1.2 as an S-system. The original
system is characterized by the two matrices:
\[
   B = \left( \begin{array}{cc} p & 0 \\ 0 & q \end{array} \right) \;\: ,
   E = \left( \begin{array}{ccc} 1 & 0 & 0 \\
      \lambda_1 & a_{11} & a_{12} \\
      \lambda_2 & a_{21} & a_{22} \end{array} \right)
\]
We may wish to transform our starting system into, for example, an S-system
defined by matrix:
\[
   \hat{E} = \left( \begin{array}{ccc} 1 & 0 & 0 \\
      \mu_1 & - \sigma_{1} &     0        \\
      \mu_2 &       0      & - \sigma_{2} \end{array} \right) \;\: , \;\: \mu_i , \sigma_i > 0,
      \;\: i=1,2.
\]
According to the prescription of Theorem 3 we shall assume that
$\mid A \mid = a_{11}a_{22} - a_{12}a_{21} \neq 0$.
Additionally, the compatibility condition implies that:
\[
    \left( \begin{array}{c} \mu_1 / \sigma_1 \\ \mu_2 / \sigma_2 \end{array} \right) =
    \; - \; A^{-1} \cdot \left( \begin{array}{c} \lambda_1 \\ \lambda_2 \end{array} \right)
    \equiv \left( \begin{array}{c} \xi_1 \\ \xi_2 \end{array} \right)
\]
The quantities $\xi_1$ and $\xi_2$ must be positive in order to have a
consistent set of equations. If this is the case we may then set, for example,
$\sigma_i = 1$ and $\mu_i = \xi_i$, $i = 1, 2$. Notice that $\hat{A} = -I$
with this choice of parameters. We infer, from Theorem \ref{th7},
$\; \: C = A \cdot \hat{A}^{-1} = -A$. Consequently, our new variables are:
\begin{eqnarray*}
   y_1 & = & x_1^{- \Delta a_{22}}x_2^{ \Delta a_{12}} \\
   y_2 & = & x_1^{ \Delta a_{21}}x_2^{- \Delta a_{11}}
\end{eqnarray*}
where $ \Delta = ( \mid A \mid )^{-1}$. We shall thereby obtain a matrix $B$
of the form
\[
    \hat{B} = B \cdot C = \; - \;
    \left( \begin{array}{cc} pa_{11} & pa_{12} \\ qa_{21} & qa_{22} \end{array} \right) \;\: ;
\]
the target S-system being given by:
\begin{eqnarray*}
  \dot{y}_1 = \xi_1 y_1 - y_1^{1+pa_{11}}y_2^{pa_{12}}  \\
  \dot{y}_2 = \xi_2 y_2 - y_1^{qa_{21}}y_2^{1+qa_{22}}
\end{eqnarray*}

\subsection{Case $m > n$}

\subsubsection{Recasting theory}
The general solution of this problem requires the introduction of one or more
auxiliary new variables: in fact, without the aid of an embedding the
recasting of a GLV system into an S-system may be rigorously impossible in
some cases, as the following theorem shows:
\begin{th}
\mbox{} \\
If $m>2n$ there exists no quasimonomial transformation (\ref{5}) which leads
from a GLV system to an equivalent S-system.
\end{th}

In systems not precluded by Theorem 4,
a unified description shall proceed by reducing to the $m=n$ case
through the same embedding technique of subsection 3.2.2. The advantage of
this particular embedding is that it leads to an expanded GLV system whose
matrix $\tilde{A}$ is regular, and allows the direct application of
Theorem \ref{th7}. We shall skip a formal description and go directly
to illustrate the matter with an example.

\subsubsection{Example with $m>n$.}
We shall recast as an S-system the tumor growth model of section 3.1.4.
Once embedded, we can follow the procedure of the $m=n$ case. We write the
corresponding extended matrices of the embedded system:
\[
   \tilde{E} = \left( \begin{array}{cccc}
      1 & 0 & 0 & 0 \\
      3\omega & 3\alpha k^{1/3} & -3\alpha k^{2/3} & \alpha k \\
      -1 & 0 & 1 & 0 \\
      -1 & 0 & 0 & 1
   \end{array} \right) \; ; \;\:
   \tilde{B} = \left( \begin{array}{ccc}
      -1/3 & 0 & 0 \\
      -2/3 & 0 & 0 \\
      -1   & 0 & 0
   \end{array} \right) \;\: ,
\]
where, for convenience, we have expanded the original vector $\lambda =
(3\omega)$ with two elements of value -1. We now investigate the recasting
of this GLV system as a target S-system of extended matrix:
\[
   \hat{E} = \left( \begin{array}{cccc}
      1 & 0 & 0 & 0 \\
      -a' & a & 0 & 0 \\
      -b' & 0 & b & 0 \\
      -c' & 0 & 0 & c
   \end{array} \right)
\]
All constants are taken as strictly positive. We shall look once more
for the fulfillment of the compatibility condition of Theorem 3, which
leads us to the matrix identity:
\[
   \left( \begin{array}{c}
      3\omega \\ -1 \\ -1
   \end{array} \right) =
   \left( \begin{array}{c}
      -3\alpha k^{1/3} a' / a + 3\alpha k^{2/3} b'/b - \alpha k c'/c \\
      -b'/b \\
      -c'/c
   \end{array} \right)
\]
We may choose, for example, the following values for our parameters:
\[
     a = b = c = b' = c' = 1 \; ; \;\: a' = \gamma \; , \;\:
     \gamma = k^{1/3} -\frac{1}{3} k^{2/3} - \alpha ^{-1} \omega k^{-1/3} \;\; ,
\]
which yields for our target S-system the following extended matrix:
\[
   \hat{E} = \left( \begin{array}{cccc}
      1 & 0 & 0 & 0 \\
      -\gamma & 1 & 0 & 0 \\
      -1      & 0 & 1 & 0 \\
      -1      & 0 & 0 & 1
   \end{array} \right)
\]
It is worth noticing how all the complexity of the system has been conveyed
into a single constant, $\gamma$. Also, from Theorem \ref{th7}, the transformation
matrix is given by $C = \tilde{A} \cdot \hat{A}^{-1} = \tilde{A}$. Since
$\hat{A} = I$, our target system is also, by definition, the single
quasimonomial canonical form. Solving for $\hat{B} = \tilde{B} \cdot C =
\tilde{B} \cdot \tilde{A}$ we obtain that the final system of equations is:
\begin{eqnarray*}
     \dot{y}_1 & = & -\gamma y_1 + y_1^{1+\mu_1} y_2^{\mu_2} y_3^{\mu_3} \\
     \dot{y}_2 & = & -y_2 + y_1^{2\mu_1} y_2^{1+2\mu_2} y_3^{2\mu_3} \\
     \dot{y}_3 & = & -y_3 + y_1^{3\mu_1} y_2^{3\mu_2} y_3^{1+3\mu_3} \; ,
\end{eqnarray*}
where the $\mu$'s are defined as in the example of section 3.1.4.
This system, which coincides with the single quasimonomial canonical form,
will be also an S-system provided that the values of the constants $k$,
$\alpha$ and $\omega$ are such that $\gamma$ is positive or zero. Since $C =
\tilde{A}$, the value of $V$ can be now straightforwardly retrieved as $V =
y_1^{-3\mu_1}y_2^{-3\mu_2}y_3^{-3\mu_3}$.

\subsection{Case $m<n$}

\subsubsection{Recasting theory}

Once again, the situation here is formally similar to the one where $m=n$, the
only differences arising from the fact that here the extended matrices $E$ and
$\hat{E}$ are not square. This means that, once $\hat{E}$ has been designed,
the equation to solve is $D \cdot \hat{E} = E$, where $D$ must comply to format
(\ref{nm2}). The existence and uniqueness of $D$ will depend on the ranks
of the matrices involved.

\subsubsection{Example with $m<n$}

We here analyze the example of 3.1.6. For the matter of sake, we choose the following
S--system extended matrix.
\[ \hat{E} = \left( \begin{array}{ccc}
            1 & 0 & 0 \\ 1 & -1 & 0 \\ 0 & 2 & -1 \\ 0 & 1 & 0 \end{array} \right) \]
Then we are led to the following matrix equation:
\[ \left( \begin{array}{cccc}
      1 & 0 & 0 & 0 \\
      0 & c_{11} & c_{12} & c_{13} \\
      0 & c_{21} & c_{22} & c_{23} \\
      0 & c_{31} & c_{32} & c_{33}
   \end{array} \right)
   \left( \begin{array}{ccc}
      1  & 0  &  0 \\
      1  & -1 &  0 \\
      0  & 2  & -1 \\
      0  &  1 &  0
   \end{array} \right) \; = \;
   \left( \begin{array}{ccc}
      1  & 0  &  0 \\
      -1 & 1  &  1 \\
      3  & 2  &  1 \\
      2  & -1 &  2
   \end{array} \right) \]
In this example there exists a unique solution, that given by
\[ C =  \left( \begin{array}{ccc}
      -1 & -1 &  2 \\
       3 & -1 &  7 \\
       2 & -2 &  5 \\
   \end{array} \right) \]
Solving for $\hat{B}$ we reach the following target S--system:
\begin{eqnarray*}
\dot{y}_{1} & = & y_{1} - y_{1}^{5}y_{2}^{-4}y_{3}^{14} \\
\dot{y}_{2} & = & 2y_{1}^{4}y_{2}^{-3}y_{3}^{14} - y_{1}^{5}y_{2}^{-6}y_{3}^{21} \\
\dot{y}_{3} & = & y_{1}^{4}y_{2}^{-4}y_{3}^{15}
\end{eqnarray*}

\section{General Nonlinear Systems within the GLV formalism}
\subsection{The auxiliary variables procedure}

In order to briefly describe the auxiliary variable method which
we have mentioned earlier, we shall proceed by a very general approach.
We may start by assuming that we have at hand an $n$-dimensional system
dependent on what we can define to be a set of $r$ functions, $\{ f_k \}$,
which do not comply to the quasimonomial
form, given by (\ref{4}). We shall elaborate on
a system which has previously rearranged into the following form:
\begin{equation}
   \dot{x}_{i} = x_{i}(\lambda _{i} + \sum_{j=1}^{m}A_{ij}
    \prod_{k=1}^{n}x_{k}^{B_{jk}}
    \prod_{s=1}^{r}f_{s}^{C_{js}}) \: , \;\:
    i = 1, \ldots ,n \; ,
   \label{2}
\end{equation}
which is a GLV--like form except for the set $\{ f_k \}$. Our concern
shall be to examine the circumstances under which (\ref{2}) can be
rewritten as a GLV system.

For this purpose, we shall take into account that most of the functions which
are actually used in modelling in a biological context (elementary, most of
them) obey equations relating their partial derivatives to members of the
properly chosen set $\{ f_k \}$ itself in the following way:
\begin{equation}
   \frac{\partial f_k}{\partial x_i}
     =  \sum_{r=1}^{m'} E_{r}^{(ik)}
    \prod_{j=1}^{n}x_{j}^{G_{rj}^{(ik)}}
    \prod_{s=1}^{r}f_{s}^{H_{rs}^{(ik)}},
         i = 1, \ldots ,n, \;\: k = 1, \ldots ,r.
    \label{3}
\end{equation}
The set $\{ f_k \}$ must be appropriately defined in order to fulfill
(\ref{3}). If, for example, the circular sine--function appears in
(\ref{2}), the cosine--function will correspondingly be present in (\ref{3}).
The set shall then include both functions, the cosine--function appearing with
an exponent zero in (\ref{2}).

Thus, if the set $\{ f_k \}$ has been appropriately defined, that is, such that
(\ref{2}) and (\ref{3}) hold, it is then straightforward to obtain a GLV form
by assigning each function $f_k$ to an extra variable $y_k$, and taking
into consideration that:
\begin{equation}
\dot{y}_k = \sum_{i=1}^{n} \frac{\partial f_k }{ \partial x_i} \dot{x}_i,
\label{4ii}
\end{equation}

\subsection{Sources of ambiguity}

So far, everything seems consistent in the procedure developed in 5.1. However,
there is some degree of arbitrariness as far as equations (\ref{2})-(\ref{4ii})
are ambiguously defined. We point at two sources of ambiguity:
\begin{enumerate}
   \item {\em The set of functions $\{f_{k}\}$ is certainly not unique. Any
      other set $\{F_{i}\}$ such as}
      \begin{equation}
         F_i = \left( \prod_{k = 1}^{r} f_{k}^{q_{ik}} \right)
         \left( \prod_{j=1}^{n} x_{j}^{p_{kj}} \right) \; , \;\: i = 1,
         \ldots ,r' \;\: ,
         \label{p1}
      \end{equation}
      {\em will also comply to the general starting prescribed forms
      (\ref{2}) and (\ref{3}). This statement holds in all three cases:
      $r<r'$; $r=r'$; $r>r'$.}
   \item {\em Any chosen set of functions $\{f_{k}\}$ may be compatible with
      more than one expression for each of its partial derivatives (\ref{3}),
      and thus generate different GLV systems.}
\end{enumerate}

We are going to see that
there is a substantial difference between the two previous propositions.
While the issue raised in point 1 will be shown to be formally cleared and settled
in three subsequent general theorems, that forwarded in point 2 cannot conversely be given a general
formal foothold, and has to rely on a rather empirical knowledge from a
collection of examples, let it be infinite. This difference will result in a
consistent solution to the problem posed by the ambiguity of point 1; a task
we undertake in the present paper. On the contrary, the problem set in point
2 cannot be formally solved, though we later comment on its consequences.

We now state three theorems in connection to point 1:
\begin{th} \label{th5}

\mbox{}

Let $\{f_{k}\}$ and $\{F_{k}\}$ be two sets of a number r of $C^1$-functions
defined on the strictly positive orthant of $\Re ^n$, which we shall denote
${\rm int} \; \Re^{n}_{+}$. For every k, we assume $f_{k} > 0$ and
$F_{k} > 0$ for all $x \in {\rm int} \; \Re^{n}_{+}$.

Then, if
\begin{equation}
   F_i = \left( \prod_{j = 1}^{r} f_{j}^{q_{ij}} \right)
   \left( \prod_{k=1}^{n} x_{k}^{p_{ik}} \right) \; ,
   \label{t5}
\end{equation}
we have:
\begin{description}
 \item[(i)] The inverse transformation from the set $\{F_k \}$ to the set $\{f_k\}$
 is defined iff det Q $\neq 0,$ with $(Q)_{ij} = q_{ij}, \;\: i , j = 1, \ldots , r.$
 \item[(ii)] The transformations defined by (\ref{t5}) constitute a parametric
   group, $\Xi \equiv \{ \xi(Q,P)\}$, where $(P)_{ik} = p_{ik}\;, \;\: k = 1 ,
   \ldots , n.$
\end{description}
\end{th}

\mbox{}

\begin{th} \label{th6}

\mbox{}

Let $r = {\rm card} \{f_k\}$ and $r' = {\rm card} \{F_i\}$ in (\ref{p1}). Suppose $r
\neq r'$ and $\rho = {\rm max} \{r , r' \}$. Let Q be an $r' \times r$ matrix
defined as in Theorem \ref{th5}. If {\rm rank}(Q) is maximum, there exists two
new sets of $\rho$ functions, in which $\{f_k\}$ and $\{F_i\}$ can be embedded,
and for which the statement of Theorem \ref{th5} holds.
\end{th}

\mbox{}

\begin{th}

\mbox{}

Equations (\ref{2}) and (\ref{3}) are form invariant under the group $\Xi$.
\end{th}

The set whose elements are themselves those sets of functions generated
through the action of the
group $\Xi$ on the set $\{f_{k}\}$  constitute a class of equivalence
$\Gamma \{f_{k}\}$. According to the procedure of subsection 5.1, each member
of a class $\Gamma$ will be mapped onto a GLV system, which will eventually
differ from that obtained by applying the same procedure to any other
element of the class. The question is to know if, in spite of that, all
elements of a class $\Gamma$ are mapped into a single GLV class of
equivalence. We will answer it in sections 5.3 and 6.

\subsection{Heuristic considerations on an illustrative model system}

Before supplying the reader with the formal theorems which will demonstrate
that the class of equivalence $\Gamma \{f_k \}$, generated by the group of
transformations $\Xi$, is mapped into a single GLV class of equivalence by
applying the procedure of section 5.1, we will make a heuristic analysis of
a simple one-dimensional model. It was introduced by Ludwig {\em et al.\/} \cite{die},
in order to simulate the evolution of the
population of the spruce budworm in the presence of predation by birds.
In dimensionless form it is:
\begin{equation}
\dot x = r x (1 - \frac{x}{k}) - \frac{x^2}{ 1 + x^2}. \label{9ii}
\end{equation}

Let us thus consider the model system defined by (\ref{9ii}). In order to
recast it into the GLV format we can initially choose the following function
$f = (1+x^2)^{-1}$ with derivative $ f'(x) = -2xf^2 $. According to the
casuistry of the previous section we shall examine what happens for a general
transform $y = F = x^p f^q , q \neq 0$. After elementary calculations, we
obtain from (\ref{9ii}):
\begin{eqnarray}
\dot{x} & = & x \left[ r - \frac{r}{k} x -x^{1-p/q}y^{1/q} \right] \nonumber \\
\dot{y} & = & y \left[ pr - \frac{pr}{k} x - p x^{1-p/q}y^{1/q} \right. \nonumber \\
 &  &  \left. -2rq x^{2-p/q}y^{1/q} + \frac{2rq}{k} x^{3-p/q}y^{1/q} +
   2q x^{3-2p/q}y^{2/q} \right] \label{26ii}
\end{eqnarray}
It is straightforward to check that the products $B \cdot A$ and $B \cdot
\lambda$ are independent of exponents $p$ and $q$. That means that all
parameter-dependent GLV systems (\ref{26ii}) do actually belong to a single
GLV class of equivalence, and makes the result of the procedure independent
of any specific choice of auxiliary variables within the class of functions
$x^pf^q$. Thus, the class $\Gamma \{f\}$ is mapped into a single GLV class of
equivalence. We shall generalize this assertion for any function in section 6.

The previous invariance is not so surprising. In fact, (\ref{t5}) and the
functions $x^pf^q$ are
disguised forms of quasimonomial transformations \cite{bre1}, and these
map into one another different GLV systems within a given GLV class of
equivalence, leaving
$B \cdot A$ and $B \cdot \lambda$ invariant. The quasimonomials of the GLV
systems are also invariants of the class \cite{bhyvf}. In the present case,
from (\ref{26ii}): $ x \; ; \; x^{1-p/q}y^{1/q} = x/(1+x^2) \; ; \;
x^{2-p/q}y^{1/q} = x^2/(1+x^2) \; ; \; x^{3-p/q}y^{1/q} = x^3/(1+x^2) \; ; \;
x^{3-2p/q}y^{2/q} = x^3/(1+x^2)^2 \; .$

We now examine the context of Theorem \ref{th6}. For that we may start from
two different sets of functions to deal with the problem, for example:
\[ \{ f_1 \} = \left\{ \frac{1}{1+x^2} \right\} \; , \;\:
   \{ F_1 , F_2 \} = \left\{ \frac{1}{1+x^2} , \frac{x^3}{1+x^2} \right\} \]
According to Theorem \ref{th6},
set $\{ f_1 \} $ is embedded into set $\{ f_1 , f_2\} $, with $f_2 \equiv 1$.
Then both sets, $\{ f_k \} $ and $\{ F_k \} $, are related through an
invertible transformation of form (\t5) with matrices
\[   Q = \left( \begin{array}{cc}
         1   &   0   \\
         1   &   1
         \end{array} \right) \;\; , \;\;\:
     P = \left( \begin{array}{c}  0 \\ 3 \end{array} \right) \]
Note that the second column of matrix $Q$ is arbitrary: the only requirement
imposed by Theorem \ref{th6} is that $Q$ must be invertible.
If we use the set $\{f_1 , f_2\}$ to carry out the substitution in (\ref{9ii}),
we arrive at the following system:
\begin{eqnarray}
\dot{x}   & = & x \left[ r - \frac{r}{k} x -xy_{1} \right] \nonumber \\
\dot{y}_1 & = & y_1 \left[ -2r x^{2}y_1 + \frac{2r}{k} x^3 y_1 + 2 x^3 y_{1}^{2} \right] \nonumber \\
\dot{y}_2 & = & 0  \label{c1}
\end{eqnarray}
If, on the contrary, we start from $\{F_1 , F_2\}$, we are led to the system:
\begin{eqnarray}
\dot{x}   & = & x \left[ r - \frac{r}{k} x -xy_{1} \right] \nonumber \\
\dot{y}_1 & = & y_1 \left[ -2r x^{2}y_1 + \frac{2r}{k} y_2 + 2 x^3 y_{1}^{2} \right] \nonumber \\
\dot{y}_2 & = & y_2 \left[ 3r - \frac{3r}{k} x -3xy_{1}-2r x^{2}y_1 + \frac{2r}{k} y_2 + 2 x^3
      y_{1}^{2} \right] \label{c2}
\end{eqnarray}
It can be easily checked that, after the apparently naive introduction of the
function $f_2 = 1$, systems (\ref{c1}) and (\ref{c2}) belong to the same class
of equivalence: both possess the same quasimonomials and the same matrix
invariants $B \cdot A$ and $B \cdot \lambda$.

We can however skip the application of Theorem \ref{th6}, and introduce both
sets of auxiliary functions independently, namely $\{f_1 \}$ and $\{F_1 ,
F_2\}$. In this case the result is the same as before, with the only
difference that the last equation in system (\ref{c1}) does not exist now.
Consequently, the two systems cannot be in a same class
of equivalence, since the number of variables is different in each case.
Nevertheless, it can be easily seen that the quasimonomials and the
matrix invariants still coincide. As a consequence, we can make use of a
general procedure we have seen in subsection 3.1.3
for the reduction of these GLV systems to a common class of equivalence:
the Lotka-Volterra embedding. When we perform such an embedding over both
systems, the result will be GLV systems with five variables and five
quasimonomials, both in the same class of equivalence. In particular, the
systems are equivalent to a $5\times 5$ Lotka-Volterra system of
matrices $\tilde{A} = B \cdot A$ and $\tilde{\lambda} = B \cdot \lambda$.
We can thus infer that, independently of the number of auxiliary variables of
form $x^p f^q$ that we introduce in system (\ref{9ii}), once the definitions
of $f$ and
its derivative are fixed, all the $(p,q)$-dependent GLV systems we obtain can
be embedded into the same class of equivalence.

We conclude the section by examining what happens when we start from
different forms of the derivative. If, for example, we set
\begin{equation}  f = \frac{x^2}{1+x^2} \end{equation}
there are in fact infinite possible definitions of the derivative for this
function, namely:
\[
   \frac{\mbox{d}f}{\mbox{d}x} = 2 x^{-2n-3}f^{n+2}(1 + x^2)^n \;\: , \;\;\:
   n = 0 , 1 , 2 \ldots
\]
It can be easily checked that, in general, different expressions of the
derivative lead to different quasimonomials, both in number and form, and
consequently to different classes of equivalence (point {\em (iii) \/} of
Theorem \ref{thg}).

\section{Embedding into the GLV form}
We shall now proceed to formalize the results obtained in section 5.3.
There will be no conceptual objection in dealing
with a single non-quasimonomial function $f$. This is
what we shall do from now on, in order to develop the essential features of
the problem with the greatest simplicity. We shall thus consider a system of
the general form:
   \[ \dot{x}_{s} = \sum_{i_{s1}, \ldots ,i_{sn},j_{s}} a_{i_{s1} \ldots
        i_{sn} j_{s}} x_{1}^{i_{s1}} \ldots x_{n}^{i_{sn}}f(\bar{x})^{j_{s}} \]
   \begin{equation}
         x_{s}(t_{0}) = x_{s}^{0}, \; \: s=1, \ldots , n 
     \label{eq:ini}
   \end{equation}

We additionally assume that $f(\bar{x})$ is such that its partial
derivatives can be expressed in the following form:
\begin{equation}
   \frac{\partial f}{\partial x_{s}} = \sum_{e_{s1},\ldots ,e_{sn},e_{s}}
      b_{e_{s1} \ldots e_{sn} e_{s}}x_{1}^{e_{s1}} \ldots x_{n}^{e_{sn}}f(\bar{x})^{e_{s}}
   \label{deriv}
\end{equation}
All constants in (\ref{eq:ini}) and (\ref{deriv}) are assumed to be
real numbers.

The procedure to transform (\ref{eq:ini}) and (\ref{deriv}) into a GLV system is
then straightforward. We know from Section 5 that this can be carried
out by introducing a set of $l$ additional variables of the form
\begin{equation}
   y_{r} = f^{q_{r}}\prod_{s=1}^{n} x_{s}^{p_{rs}}, \:\; q_{r} \neq 0, \; \;
   \forall \; r = 1 \ldots l \; ,
   \label{eq:cam}
\end{equation}
with real exponents $q_{r}, \: p_{rs}$. For the time being, we shall
assume that a given value of $l$ is selected, that is, we shall deal with a
fixed number of auxiliary variables. We will later release this requirement.

The introduction of the auxiliary variables (\ref{eq:cam}) leads to the
following system for the original variables:
\begin{equation}
   \dot{x}_{s} = x_{s} \left[ \sum_{i_{s1}, \ldots ,i_{sn},j_{s}} a_{i_{s1} \dots
      i_{sn} j_{s}}y_{1}^{j_{s}/q_{1}}\prod_{k=1}^{n} x_{k}^{i_{sk}- \delta _{sk} -
      j_{s}p_{1k}/q_{1}} \right]
   \label{eq:x}
\end{equation}
for $s = 1, \ldots ,n$. As usual, $ \delta _{sk} = 1$ if $s=k$, and 0
otherwise. For the new variables (\ref{eq:cam}) we obtain
\[
   \dot{y_{r}} = \sum_{s=1}^{n} \frac{\partial y_{r}}{\partial x_{s}}
         \dot{x} _{s} =
   y_{r} \left[ \sum_{s=1}^{n} \{ p_{rs}x_{s}^{-1}\dot{x}_{s} + \right. \]

\begin{equation}
   \left. + \sum_{i_{s \alpha },j_{s},e_{s \alpha },e_{s}}a_{i_{s \alpha },j_{s}}
   b_{e_{s \alpha }
   e_{s}} q_{r} y_{r}^{(e_{s}+j_{s}-1)/q_{r}}\prod_{k=1}^{n}x_{k}^{i_{sk}+e_{sk}
   + (1-e_{s}-j_{s})p_{rk}/q_{r}} \} \right]
   \label{eq:y}
\end{equation}
where $\alpha = 1, \ldots ,n $. Appropriate initial conditions $y_{r}(0)$
must also be included (this will be assumed whenever a new variable is
introduced). Thus, with (\ref{eq:x}) and (\ref{eq:y}) the reduction of system
(\ref{eq:ini}) to the GLV format is achieved. Notice that, from
(\ref{eq:cam}), the expression of $f$ in terms of the $y_{r}$ is not unique.
It has been specified, in (\ref{eq:x}), in terms of $y_1$, but this could have been
done by choosing any other variable $y_r$. We will prove that this choice is
irrelevant.

Let us now focus attention on the generic GLV system
(\ref{eq:x})--(\ref{eq:y}). It is clear that different systems are obtained
for distinct
choices of the auxiliary variables (\ref{eq:cam}). We shall first demonstrate
that all these systems are part of one and the same equivalence class
\cite{bre1,bhyvf}, that is:
\begin{th}     \label{th8}
 \mbox{} \\
  Let us assume a specific realization for equations (\ref{eq:ini}) and
  (\ref{deriv}). Then, all GLV systems -eqs. (\ref{eq:x})--(\ref{eq:y})-
  generated by the introduction of a given number $l$ of auxiliary
  variables (\ref{eq:cam}) belong to the same class of equivalence.
\end{th}

All systems complying with the format (\ref{eq:x})--(\ref{eq:y}) are in the 
same GLV class of equivalence: according to the previous results they must 
thus possess identical quasimonomials. This can be easily 
checked if we rewrite such quasimonomials in terms of the original variables
$\bar{x}$ and $f(\bar{x})$. The corresponding equations for the $x_{s}$ are
\begin{equation}
   \dot{x}_{s} = x_{s} \left[ \sum_{i_{s1}, \ldots ,i_{sn},j_{s}} a_{i_{s1} \dots
      i_{sn}j_{s}}f^{j_{s}}\prod_{k=1}^{n} x_{k}^{i_{sk}-\delta _{sk}} \right]
   \label{eq:xf}
\end{equation}
with $s=1, \ldots ,n$. For the $y_{r}$ we obtain:
\begin{equation}
   \dot{y}_{r} = y_{r} \left[ \sum_{s=1}^{n} \{ p_{rs}x_{s}^{-1}\dot{x}_{s} +
   \sum_{i_{s \alpha }j_{s}e_{s \alpha }e_{s}}a_{i_{s \alpha }j_{s}}b_{e_{s
   \alpha }e_{s}}q_{r}f^{e_{s}+j_{s}-1}\prod_{k=1}^{n}x_{k}^{i_{sk}+e_{sk}} \} \right]
   \label{eq:yf}
\end{equation}
where $\alpha = 1, \ldots ,n$. The quasimonomials, as functions of
$x_k , \;\: k=1, \ldots, n$,
do not depend in any way on the definition of the auxiliary variables
(\ref{eq:cam}), but only on constants from (\ref{eq:ini})--(\ref{deriv}).

We can then state the following proposition:
\begin{co}
 \mbox{} \\
 For a given number l of auxiliary variables, the GLV class of equivalence
 in which (\ref{eq:ini}) is embedded is completely determined by the
 choices for $f(\bar{x})$ and the particular representation of its
 derivatives (\ref{deriv}).
\end{co}

Irrespectively of the number $l$ of auxiliary variables, we have classes of
equivalence with the same $m$ quasimonomials. Moreover, in all
these classes the matrix invariants $B \cdot A$ and $B \cdot \lambda$ are
of sizes $m \times m$ and $m \times 1$, respectively. This makes us 
suspect that those products can also be identical, as in the case of the
quasimonomials. This would imply that, after an appropriate embedding (see
subsection 3.1.3), we could include all those systems into
the same class of equivalence of $m$-dimensional systems, independently of
the actual value of $l$. The next theorem shows that this is the case.

\begin{th}  \label{th10}
 \mbox{} \\
Let us assume, as in Theorem \ref{th8}, a specific realization for eqs.
(\ref{eq:ini}) and (\ref{deriv}).
Let us consider the set $\Phi$ of all (n+l)-dimensional systems,
(\ref{eq:x})-(\ref{eq:y}), obtained from all possible choices of
auxiliary variables (\ref{eq:cam}), as l varies
in the interval $1 \leq l \leq m-n$, where m is the number of quasimonomials.
Then, all elements of $\Phi$ can be embedded in a single GLV class of
equivalence of m-dimensional systems.
\end{th}

Thus it has been demonstrated that, independently of the number $l, \;
1 \leq l \leq (m-n)$ and specific form of variables of type (\ref{eq:cam}),
the final class obtained is always the same,
since the $m$ invariant quasimonomials and the matrices $B \cdot A$ and
$B \cdot \lambda$ are completely independent of those degrees of freedom.
We also have the following:

\begin{co}   \label{co11}
\mbox{} \\
All elements of the set $\Phi$ can be embedded in a unique m-dimensional
Lotka-Volterra system.
\end{co}

\section{Direct calculation of the quasimonomials}

An important issue in the practical application of Theorem \ref{th10} and
Corollary \ref{co11} is that of the knowledge of the $m$ quasimonomials
which are actually going to be the variables of the target $m$-dimensional
Lotka-Volterra system. These quasimonomials can be evaluated directly from
equations (\ref{eq:ini})-(\ref{deriv}),
without any need to first comply with the GLV form (\ref{eq:x})-(\ref{eq:y}).

An important previous definition is that of {\em vector of exponents\/}
associated with 
a given quasimonomial $x_{1}^{i_{1}} \ldots x_{n}^{i_{n}}f(\bar{x})^{j}$. The
vector of exponents is a shorthand notation in which this quasimonomial is
expressed as $(i_{1}, \ldots ,i_{n} | j )$. Since the order of the variables
is implicit in this ``vector'', all the information about the quasimonomial
is contained in it.

Thus the algorithm consists of the following sequential rules:
\begin{enumerate}
   \item Build the vectors of exponents for all the quasimonomials present in
      both equations (\ref{eq:ini}) and (\ref{deriv}). We shall label such
      sets of vectors $E_{s}$ for the $s$-th equation in (\ref{eq:ini}):
\[
      E_{s} = \{ (i_{s1}, \ldots ,i_{sn} \mid j_{s}) \} \;\: ,
\]
      and $D_{s}$ for the $s$-th equation in (\ref{deriv}):
\[
      D_{s} = \{ (e_{s1}, \ldots ,e_{sn} \mid e_{s}) \} \;\: .
\]
      Notice that these families of vectors must be constructed separately for
      each one of equations (\ref{eq:ini}) and (\ref{deriv}).
   \item In the $\dot{x} _{s}$ equation of the GLV system the quasimonomials
      will be:
      \[ \{ (i_{s1}, \ldots ,i_{ss}-1, \ldots ,i_{sn} \mid j_{s}) \} \]
      This expression is obviously to be applied over all vectors in $E_{s}$.
   \item While in the $\dot{y} _{r}$ GLV equation:
      \begin{itemize}
         \item Those quasimonomials already present in $\dot{x} _{s}$, for all
            $s = 1, \ldots ,n$ such that $p_{rs} \neq 0$.
         \item Those coming from all pairwise combinations of elements of
            $E_{s}$ and $D_{s}$, such that:
      \[ (i_{s1}+e_{s1}, \ldots ,i_{sn}+e_{sn} \mid j_{s}+e_{s}-1) \;\: , \]
            for all $s = 1, \ldots ,n$.
   \end{itemize}
\end{enumerate}

This constructive evaluation also makes straightforward the derivation of
algebraic expressions for bounds to the number $m$ of quasimonomials, or,
equivalently, upper and lower bounds to the phase-space dimension of the
target class, as we have:

\begin{th}   \label{th12}
\mbox{} \\
If $\epsilon _{s} = {\rm card}(E_{s})$ and $\delta _{s} =
{\rm card}(D_{s})$ :
\begin{itemize}
   \item An upper bound to $m$ is:
   \[ \omega = \sum_{s=1}^{n}(1+\delta _{s})\epsilon _{s} \]
   \item A lower bound to $m$ is:
   \[ \alpha = {\rm card}\{ \bigcup_{s=1}^{n}[E_{s}\backslash (\delta _{s1},
      \ldots ,\delta _{sn} \mid 0)] \} + \delta _{1n} \;\: , \]
\end{itemize}
where $\delta _{\alpha \beta }$ is Kronecker's delta.
\end{th}

We can observe that,
the shorter the expressions for the derivatives, the smaller the upper bound.
This establishes an ecomomy principle in election (\ref{deriv}).

To illustrate the preceding results, we can make use of the example of
section 5.3:
\[ \dot x = r x (1 - \frac{x}{k}) - \frac{x^2}{ 1 + x^2} \; , \]
\[ f = \frac{1}{1+x^2} \;\: , \;\;\: \frac{\mbox{d}f}{\mbox{d}x} = -2xf^2 \]
We shall evaluate the quasimonomials resulting from the introduction of a
generic auxiliary variable of the form $y = x^p f^q$, $q \neq 0$. The sets
of vectors are $ E_1 = \{ ( 1 \mid 0 ) , ( 2 \mid 0 ) , ( 2 \mid 1 ) \} $,
and $D_1 = \{ ( 1 \mid 2 ) \} $.
We thus have $\epsilon = 3$ and $\delta = 1$. Since card$\{ E_1 \backslash
( 1 \mid 0 ) \} \; = \: 2$ and $n = 1$, the bounds to the number of
quasimonomials are $\alpha = 3$ and $\omega = 6$.

By applying the second rule, we see that the quasimonomials in the $\dot{x}$
equation are: $( 0 \mid 0 )$, which is a constant and must be discarded;
$( 1 \mid 0 ) = x$ and $( 1 \mid 1 ) = xf$. From the third rule,
if $p \neq 0$, the set $\{ x , xf \}$ will also be present in the $\dot{y}$
equation. This equation will always contain the quasimonomials $( 2 \mid 1 )
= x^2 f$, $( 3 \mid 1 ) = x^3 f$ and $( 3 \mid 2 ) = x^3 f^2$.

Thus, we find five different quasimonomials. The confirmation that they
coincide with those in the explicit system (\ref{26ii}) is straightforward
(see Section 5.3).

\section{Final comments}
The structure of the generalized Lotka-Volterra equations has been shown to
provide an ideal setting for codifying in a unified framework the relations
between different formats of practical use when modelling in terms of
ordinary differential equations.

When encapsulated in the context of the GLV formalism, a given model finds
itself embedded in a class of equivalence, the members of which define an
infinite set of models topologically equivalent to one another. It is found
that a class of equivalence may contain several members belonging to those
representative canonical forms we mentioned before, each one of them being
mapped into another by the quasimonomial transformation rules, but a single
LV representative.
For these reasons, together with the Hamiltonian properties of the
Lotka-Volterra equations \cite{pla}, this form may rightly deserve to be
called {\em canonical.\/}

Finally, if the GLV formalism is to be of any practical use, additionally to
its algebraic structure it must comply to the requirement of providing a
framework for the fairly general class of model systems whose vector field
either does not obey its format or cannot be rewritten in some trivial way
in terms of it. We have demonstrated that this is possible. The way in which
this can be unambiguosly done has been the issue of the concluding sections
of the work.

\mbox{}

{\em Acknowledgements: This work has been supported by the DGICYT (Spain),
under grant PB94-0390. B. H. acknowledges a doctoral fellowship from
Comunidad Aut\'{o}noma de Madrid. The authors also thank Drs. A. Sorribas
and E. O. Voit for supplying them with copies of their articles.}

\pagebreak
{\Large\bf Appendix}

\mbox{}

We carry out those demonstrations of previous theorems which are nontrivial.

\mbox{}

{\em Proof of Theorem \ref{th8}.}

Given two different choices of auxiliary variables
\[ y_{r} = f^{q_{r}}\prod_{s=1}^{n} x_{s}^{p_{rs}} , \;\:
   y'_{r} = f^{q'_{r}}\prod_{s=1}^{n} x_{s}^{p'_{rs}} ,
   \:\; r = 1, \ldots ,l ; \; q_{r} , q'_{r} \neq 0 \; \forall r ,
   \]
the resulting sets of GLV variables will be, respectively,
$(x_{1}, \ldots ,x_{n}, y_{1} , \ldots , y_{l}) \equiv (x_{1}, \ldots ,
x_{n+l})$ and $(x_{1}, \ldots ,x_{n},
y'_{1}, \ldots , y'_{l}) \equiv (\hat{x}_{1}, \ldots ,\hat{x}_{n+l})$.
An easy calculation shows that both sets of variables are connected
through a quasimonomial transformation of the kind
\[ x_{i} = \prod_{k=1}^{n+l} \hat{x}_{k}^{C_{ik}} , \;\: i = 1 , \ldots ,
           n + l \;\: ,\]
with $C$ given by:
\[ C = \left( \begin{array}{cc}
        I_{n \times n}  & O_{n \times l}  \\
        \tilde {A}_{l \times n} & \tilde {B}_{l \times l}
        \end{array} \right) \;\: , \]
where the dimension of the submatrices is indicated by the subindexes, $I$ is
the identity matrix, $O$ is the zero matrix, $\tilde {A}_{ij} = \alpha_{ij}$,
$\tilde {B}_{ij} = \beta_{i} \delta_{ij}$, with
\[ \alpha_{rs} = p_{rs} - \beta_{r} p'_{rs} , \;\: \beta_{r} =
\frac{q_{r}}{q'_{r}} \; . \]
Consequently, both systems are members of the same class.$\Box$

\mbox{}

{\em Proof of Theorem \ref{th10}.}

We shall write for simplicity the GLV system
(\ref{eq:x})-(\ref{eq:y}) in the equivalent form
\[
   \dot{x_{i}} = x_{i}(\lambda _{i} + \sum_{j=1}^{m}A_{ij}\prod_{k=
      1}^{n+l}x_{k}^{B_{jk}}) , \;\: i = 1, \ldots ,n+l \;\: , \; m \geq n+l.
\]
When we perform the changes (\ref{eq:cam}) for different numbers of new
variables, that is, for different values of $l$, we arrive at different-sized
GLV systems. However, for a given $l$ all of them lead to the same
Lotka-Volterra system, since they belong to a same GLV class of equivalence.
Moreover, from (\ref{eq:x})-(\ref{eq:y}) we know that all of them possess
the same quasimonomials (the Lotka-Volterra variables) independently
of $l$. Thus, in order to complete this proof, we only need to
demonstrate that the Lotka-Volterra matrices $B' = B \cdot A$ and $\lambda '
= B \cdot \lambda$
do not depend on $l$, either. We can do this by induction.

First, we consider the case $l=1$. If we introduce a single new variable
$y_{1}=x_{n+1}$ of the kind (\ref{eq:cam}), the GLV system will be
characterized by three matrices $A, B, \lambda$. If we add to this system
a second auxiliary variable $x_{n+2}=y_{2}$, a simple way to write the
corresponding $m \times (n+2)$ matrix $\tilde{B}$ is:
\[   \tilde{B} = (B,\vec{0})  \]
where the $\vec{0}$ indicates that the last column of $\tilde{B}$ is composed
of zeros. We shall denote as $\tilde{A}$ and $\tilde{\lambda}$ the other two
matrices of this system for which \mbox{$l=2$}. Note that $\mbox{rank}(\tilde{B})=
n+1$, so this matrix is not of maximal rank and cannot be employed to
perform an embedding of the kind showed in subsection 3.1.3. Nevertheless,
this will not affect this demonstration, since we shall consider all possible
matrices, including those of maximal rank, starting from this one:
suppose that we reexpress the quasimonomials as powers of the variables
$x_1 , \ldots , x_n , x_{n+1} , x_{n+2}$, but with a choice of the exponents
different to the one given by $\tilde{B}$. These exponents will be the
entries of a new matrix $\tilde{B}'$ (there are infinite possibilities for
this alternative description).
The equality of the i-th quasimonomial, $1 \leq i \leq m$ implies:
\begin{equation}
   x_{1}^{b_{i,1}} x_{2}^{b_{i,2}} \ldots x_{n+1}^{b_{i,n+1}} =
      x_{1}^{b_{i,1}'} \ldots x_{n}^{b_{i,n}'} x_{n+1}^{b_{i,n+1}'}
      x_{n+2}^{b_{i,n+2}'} \; ,
   \label{eq:0}
\end{equation}
where $b, b'$ are the elements of $\tilde{B}, \tilde{B}'$ respectively.
It is important to notice that $x_{n+2}$ is not independent of the rest of
the variables $x_1 , \ldots , x_n , x_{n+1} $:
\[ x_{n+2} = x_{n+1}^{\alpha}\prod_{k=1}^{n}x_{k}^{\beta_{k}}, \;\:
      \alpha \neq 0 \; . \]
If we substitute this expression into (\ref{eq:0}) and regroup in the left
side we are led to:
\begin{equation}
   \left(\prod_{j=1}^{n}x_{j}^{b_{i,j}-b'_{i,j}-\beta_{j}b'_{i,n+2}}\right)x_{n+1}^{b_{i,n+
      1}-b'_{i,n+1}-\alpha b'_{i,n+2}} = 1 \; .
   \label{eq:1}
\end{equation}
Since all the variables in (\ref{eq:1}) are independent, the only solution is
that the exponents are identically zero. We can rewrite this condition as:
\[ \tilde{B}' = \tilde{B} - Q \;\: , \]
with
\[
   (Q)_{ij} = \left\{ \begin{array}{c} b_{i,n+2}' \beta_j \; , \;\: 1 \leq j \leq n \\
              b_{i,n+2}' \alpha \; , \;\: j = n + 1 \\
              - b_{i,n+2}' \; , \;\: j = n + 2 \end{array} \right.
\]
The relationships among the coefficients of the quasimonomials in
(\ref{eq:x})--(\ref{eq:y}) imply the vanishing of the products
$Q \cdot \tilde{A}$ and $Q \cdot \tilde{\lambda}$. This step of the proof
involves a simple but lengthy algebra. We do not detail it.
Then, $\tilde{B}' \cdot \tilde{A} = \tilde{B} \cdot \tilde{A}$, and
$\tilde{B}' \cdot \tilde{\lambda} = \tilde{B} \cdot \tilde{\lambda}$, and the
theorem is proved for the case $l=2$.

In a similar way it can be shown that this equality holds for the step
$l \rightarrow l+1$ in the induction procedure. Again the algebra is simple
but the process is tedious, and will not be reproduced here.$\Box$

\mbox{}

{\em Proof of theorem \ref{th12}.}

The lower bound in the case $n \neq 1$ and the upper bound are obvious
consequences of rules 2 and 3 of the algorithm. The
demonstration then reduces to prove that if $n = 1$
\[ \omega = \mbox{card}\{ E_{1} \backslash (1 | 0) \} + 1  \;\: . \]
To understand the addition of 1 to this expression we need a previous lemma:

\begin{flushleft}
{\em LEMMA }
\end{flushleft}
{\em If $n=1$ there exists $( \alpha | \beta ) \in D_{1}$ such that
$( \alpha | \beta ) \neq ( -1 | 1 )$. }

\mbox{}

{\em Proof of the Lemma.}

Suppose that the lemma is false: then $D_{1} = \{ (-1 | 1) \} $. Consequently,
\[  \frac{\mbox{d}f}{\mbox{d}x} = kx^{-1}f     \]
and $f$ is of the form $cx^{k}$ with $c$ a real constant. This contradicts
the assumption of a nonquasipolynomial $f$ and demonstrates the lemma.$\Box$

We return to the Theorem.
Taking into account the fact stated in this lemma, the proof for the case
$n=1$ holds by a simple counting procedure: if $D_{1} = \{ (-1 | 1) \} $,
then rule 3 of the algorithm automatically repeats the quasimonomials
generated from rule 2, all of them of the form $ (i-1 | j) $. However, since
$D_{1} \neq \{ (-1 | 1) \} $, at least one new quasimonomial will appear from
rule 3 which is not generated by rule 2.$\Box$

\pagebreak

\end{document}